\documentclass[utf8]{FrontiersinHarvard}
\usepackage{url,hyperref,lineno,microtype,subcaption}
\usepackage[onehalfspacing]{setspace}

\def\keyFont{\fontsize{8}{11}\helveticabold }
\def\firstAuthorLast{Zeshan {et~al.}} 
\def\Authors{Muhammad Zeshan Ashraf\,$^{1}$, Wenyuan Cui\,$^{1,*}$ and Hongjie Li\,$^{2}$}

\def\Address{$^{1}$College of Physics, Hebei Normal University, Shijiazhuang, 050024, China \\
$^{2}$College of Sciences, Hebei University of Science and Technology, Shijiazhuang, 050018, China}
\def\corrAuthor{Wenyuan Cui}

\def\corrEmail{cuiwenyuan@hebtu.edu.cn}

\begin{document}
\onecolumn

\title{The Origins of Neutron-capture Elements In Globular Cluster
M22} 

\author[\firstAuthorLast ]{\Authors} 
\address{}
\correspondance{} 
\extraAuth{}
\maketitle
\begin{abstract}

The chemical abundances of metal-poor (MP) stars in globular clusters provide valuable information for constraining their evolutionary scenarios. Using both main $r$-process and weak $r$-process patterns, we fit the abundances of $s$-poor stars in the globular cluster M22. The coefficients of the main and weak $r$-process components are nearly constant for the sample stars, including $s$-rich stars. By considering the contribution of the $s$-process from low-mass AGB stars, the abundances of $s$-rich stars in M22 can also be fitted effectively. Furthermore, the increasing trend in the $s$-process component coefficients ($C_s$) with increasing [Fe/H] suggests a gradual increase in the contribution from low-mass AGB stars.

\tiny
 \keyFont Keywords: Metal poor stars --- nucleosynthesis --- main r-process --- weak r-process
\end{abstract}

\section{Introduction}

Metal poor (MP) stars are typically old and are often regarded as cosmic fossils due to their abundance patterns, which preserve valuable information about the early stages of galactic evolution. The analysis of elemental abundance patterns is fundamental for understanding the chemical evolution of galaxies and refining nucleosynthesis theory, which describes the formation of elements in stars and other astrophysical environments. Elements with atomic numbers $8 \leq Z \leq 20$ are classified as light elements, primarily synthesized through fusion processes in stellar cores during their evolutionary phases. Elements with atomic numbers $21 \leq Z \leq 30$ are referred to as iron-group elements, which are predominantly produced through explosive nucleosynthesis in supernovae and represent the final products of stellar fusion in massive stars. Elements with $Z > 30$ are categorized as neutron-capture elements, which are further divided into lighter neutron-capture elements for $31 \leq Z < 56$ and heavy neutron-capture elements for $Z \geq 56$ \citep{Sneden2008}.

The synthesis of heavy elements in the cosmos is governed by two distinct mechanisms: the slow neutron-capture process ($s$-process) and the rapid neutron-capture process ($r$-process) \citep{Burbidge1957}. Observational evidence and nucleosynthesis theories suggest that the two processes occur in distinct physical conditions and astronomical environments. The $s$-process primarily takes place in the vicinity of the \(\beta\)-stability valley, as the time interval between consecutive neutron captures is significantly longer than \(\beta\)-decay. It can be further classified into the main and weak $s$-processes. The main $s$-process takes place during the asymptotic giant branch (AGB) phase of stars with medium and low masses (1.3-8$\ M_\odot$). In contrast, the weak $s$-process occurs during the core He-burning and C-shell burning phases of massive stars \citep{Busso1999}. The $r$-process often takes place in a very energetic cosmic environment. Due to the substantial absorption of neutrons by the target nuclei, the nuclei become highly unstable, posing significant challenges in investigating the source of $r$-process elements \citep{Sneden2008}. 

Observations of the ultra metal-poor halo stars CS 22892-052 \citep{Sneden2003, Cowan2005} and CS 31082-001 \citep{Hill2002, Honda2004} have revealed that their heavier elements ($Z \geq 56$) are in remarkable agreement with the abundance patterns of solar $r$-process elements \citep{Sneden2000}. The mechanism responsible for generating such an abundance pattern is termed the "main $r$-process" \citep{Truran2002, Wanajo2006}. However, the lighter neutron-capture elements ($37 \leq Z \leq 47$, i.e., from Rb to Ag) in these stars do not fully align with the solar-system $r$-process residual pattern \citep{Sneden2000, Hill2002}. This discrepancy suggests that the $r$-process abundance pattern in solar-system material cannot be explained by a single process, such as the main $r$-process.
Several potential sites have been proposed for the main $r$-process, including core-collapse supernovae and binary neutron star mergers \citep{Burbidge1957, MacFadyen1999, Winteler2012, Abbott2017}. Other suggested sources of the $r$-process include magneto-rotational supernovae and collapsars \citep{Thielemann2023}. However, the precise origin of the main $r$-process remains uncertain \citep{Sneden2008, Farouqi2022}.
On the other hand, observations of very metal-poor stars, such as HD 122563 and HD 88609 \citep{Westin2000, Johnson2002, Aoki2005, Honda2007}, show that their lighter neutron-capture elements (Sr, Y, and Zr) are in excess. Such an abundance pattern is produced by another component, referred to as the lighter element primary process (LEPP) or ''weak $r$-process''component \citep{Travaglio2004, Cowan2005, Ishimaru2005, Montes2007, Izutani2009}. This weak $r$-process explains the $r$-process abundances of the solar system for lighter neutron-capture elements. The sites of the weak $r$-process are probably Type II supernovae (SNe II) with progenitors of $M\geq\ 10M_\odot$ \citep{Travaglio2004, Ishimaru2005}. The ratios of [Eu/Fe] $\approx -0.5$ and [Sr/Fe] $\approx 0$ in HD 122563 and HD 88609 stars suggest that weak $r$-process elements are produced in conjunction with Fe and light elements rather than with heavier $r$-process elements. Based on observations of metal-poor stars with varying [Sr/Eu] ratios, \citet{Montes2007} concluded that the weak $r$-process produces a uniform and distinct abundance pattern for lighter neutron-capture elements. \citet{Zhang2010} analyzed the abundances of 12 metal-poor $r$-rich stars with metallicities [Fe/H] $< -2.1$ and found that the abundance patterns of both neutron-capture and light elements could best be explained by stars formed in a molecular cloud that had been polluted by both weak $r$- and main $r$-process material.

For many years, globular clusters have played an important role in testing many aspects of stellar evolution and stellar nucleosynthesis. In this context, the abundance patterns of neutron-capture elements offer valuable insights into stellar nucleosynthesis and the chemical evolution of globular clusters. \citet{Marino2009} performed a chemical abundance analysis of the globular cluster M22, and reported that it exhibits an intrinsic Fe abundance spread based on optical spectroscopic analyses. However, more recent work using high-resolution, homogeneously analyzed near-infrared spectra from the APOGEE survey \citet{Meszaros2020} did not confirm such a spread, concluding instead that M22 is chemically homogeneous in Fe. A key consideration is that the stellar samples analyzed in these two studies are not identical, which may partially account for the contrasting results. Differences in evolutionary stage, spatial distribution, or membership probability of the selected stars can significantly influence the derived abundance patterns. It remains possible that distinct sub-populations exist within M22, and that Fe variations, if present, are confined to a subset of stars not captured in the APOGEE sample. This highlights the importance of sample selection and motivates further targeted studies to resolve the presence or absence of Fe variations in M22. In particular, based on the abundance analysis of 35 stars, \citet{Marino2011} found that M22 exhibits a complex chemical pattern. They reported the presence of two distinct stellar groups in this cluster, characterized by significant differences in the neutron-capture elements Y, Zr, Ba, and La, namely, $s$-poor and $s$-rich groups. The presence of multiple stellar groups in M22 indicates that this cluster has undergone a complex chemical enrichment history. The stars responsible for the nucleosynthesis and the nature of the pollution mechanisms in M22 remain unknown. In the solar system, the $s$-process is attributed to two components: the main component and the weak component. However, in the globular cluster M22, the origins of neutron-capture elements, particularly $s$-process elements, are not well understood \citep{Marino2011}, and the characteristics of $s$-process nucleosynthesis remain uncertain. A quantitative understanding of the origins of neutron-capture elements in M22 has so far remained a challenging problem. Clearly, detailed studies of this cluster are essential for improving our understanding of neutron-capture processes, identifying the possible astrophysical origins, and constraining their relative contributions. These factors have motivated us to investigate the elemental abundance patterns of stars in M22, including $\alpha$-elements, iron-peak elements, and neutron-capture elements.

In this paper, we fit the abundances of 30 stars in M22 with a parametric model and calculate the relative contributions from individual neutron-capture processes to the elemental abundances in these stars. To analyze the origins of neutron-capture elements in M22, the parametric model used for the abundance decomposition of 30 stars in M22 is described in Section \ref{sec:2}. The calculations and best-fit results are presented in Section \ref{sec:3}. Our conclusions are summarized in Section \ref{sec:4}.

\section{The Origins of The Neutron-Capture Elements In M22}
\label{sec:2}
Since the main $r$-process elements are not produced in conjunction with light elements and iron-group elements \citep{Qian2007}, the neutron-capture and light-element abundance patterns of main $r$-process stars could be explained by a star formed in a molecular cloud that was initially polluted by weak $r$-process material and subsequently enriched by main $r$-process material.

\citet{Li2013} derived the main $r$- and weak $r$-components from the abundances of main $r$-process stars (CS 22892-052 and CS 31082-001; \citep{Sneden2003, Sneden2008, Hill2002}) and weak $r$-process stars (HD 122563 and HD 88609; \citep{Honda2004, Honda2007, Johnson2002}). They obtained the weak $r$-process abundance pattern by subtracting the average abundances of main $r$-process stars from the average abundances of weak $r$-process stars, normalized to Eu. This implies that all Eu is produced by the main $r$-process. They then derived the main $r$-process abundance pattern by subtracting the weak $r$-process abundance pattern from the average abundances of main $r$-process stars, normalized to Fe. This suggests that all Fe is produced alongside weak $r$-process events.

\subsection{ Parametric Model and Calculations}
The chemical elements in stars usually come from the molecular clouds where they were born, and they can be produced through multiple mechanisms. In general, the formation of elements with \(Z < 30\) is related to Supernova explosions, and the neutron-capture elements are produced by the $s$- and/or $r$-process \citep{woosley1995,roederer2010}. We start by exploring the origin of the neutron-capture elements in M22 by comparing the observed abundances with the predicted main $s$-  main $r$- and weak $r$-process contributions. For this purpose, we propose that the abundance for the \(i\)th element in a star can be calculated by the equation \citep{liang2012}:
\begin{equation}
N_i([\text{Fe/H}]) =(C_{s} N_{i,s} + C_{r,m} N_{i,r,m} + C_{r,w} N_{i,r,w}) \times 10^{[\text{Fe/H}]}, \
\label{eq:1}
\end{equation}
where \(N_{i,s}\) , \(N_{i,r,m}\) and \(N_{i,rw}\) are the abundances of the \(i\)th element produced by the main $s$-process, main $r$-process, and weak $r$ process, respectively. \(C_{s}\) , \(C_{r,m}\) and \(C_{r.w}\) are the corresponding component coefficients. Using component coefficients, we can determine the relative contributions of each process to the elemental abundances and then compare them with the corresponding component coefficients of the solar system. We adopt the component coefficients $C_s$ , \(C_{r,m}\) , and $C_{r,w}$ as 1 for the solar system standard \citep{Li2013}. This approach facilitates a better comparison of the astrophysical mechanisms contributing to the elemental abundances of our sample stars relative to the solar system. If the component coefficients for the sample stars are less than 1, it indicates that the associated astrophysical mechanisms contribute less to the sample stars than to the solar system and vice versa. These coefficients can be determined by comparing the calculated abundances with the observed abundances and minimizing the value of $\chi^2$. \(N_{i,r,m}\) and \(N_{i,r,w}\) in this model are taken from \citep{Li2013}. In order to investigate the origin of $s$-process elements , the main $s$-process abundances (\(N_{i,s}\)) in Equation (\ref{eq:1}) are taken from the calculated result of 1.5 $\, M_\odot$ AGB model at [Fe/H] = -2 with ST/12 presented by \citep{bisterzo2010}. The ST case is a standard AGB model incorporating a $^{13}$C-pocket, a localized intershell region enriched with $^{13}$C, which acts as the primary neutron source via the ${}^{13}\mathrm{C}(\alpha,\mathrm{n}){}^{16}\mathrm{O}$ reaction during the $s$-process. It was adopted by \citep{Gallino1998} and so named by later scholars. This amount of $^{13}$C for AGB stars in the 1.5–3 $M_\odot$ mass range at [Fe/H] = -0.3 appears to explain the main solar component of the s-process. The ST/12 case corresponds to a reduced  $^{13}$C pocket efficiency, which better reproduces the $s$-process abundance patterns observed in relatively metal-poor stars. The abundances of the AGB model are normalized to the main $s$-component of Ba abundance in the solar system. To determine the three coefficients in Equation (\ref{eq:1}), \(\chi^2\) is defined as:
\begin{equation}
\chi^2=\sum_{i=1}^{n}\frac{\big(logN_{i,obs}-logN_{i,cal}\big)^2}{\big(\Delta logN_{i,obs}\big)^2 \big(K-K_{free}\big)},
\label{eq:2}
\end{equation}

where \(\log N_{i,\text{obs}}\) and \(\Delta \log N_{i,\text{obs}}\) are the observed abundance and error of the \(i\)th element, which are adopted from \citet{Marino2011}. \(\log N_{i,\text{cal}}\) is the calculated abundance of the \(i\)th element, and it can be determined by Equation (\ref{eq:1}). \(K\) and \(K_{\text{free}}\) are the number of elements studied and the number of free parameters, respectively. For an optimal fit, the value of \(\chi^2\) should be close to 1 or of the order of unity.

\section{RESULTS AND DISCUSSIONS}
\label{sec:3}
We performed our calculations based on Equations (\ref{eq:1}) and (\ref{eq:2}) using the observed abundances of $\alpha$-elements (Mg, Si, Ca, Ti), Fe-peak elements (Fe, Cu, Zn), and neutron-capture elements (Y, Zr, Ba, La, Nd, Eu) in M22 stars \citep{Marino2011} and derived the component coefficients for the best fit. The best-fit results between the observed and predicted abundances are presented in Figure~\ref{fig:enter-labe1}, and the component coefficients, along with the minimum $\chi^2$ values are listed in Table~\ref{tab:1}.

To enable a rigorous comparison between the predicted nucleosynthetic yields and the observed stellar abundances, we plot the observed data as filled circles, while the solid black lines denote the best-fit model abundances derived from our component decomposition analysis. A visual inspection reveals that the theoretical predictions align remarkably well with the observed elemental abundances across all sample stars, with discrepancies generally falling within the bounds of observational uncertainty.

To quantitatively assess the quality of the fit, we present in the top panel of Figure~\ref{fig:enter-labe2} the relative offsets, defined as $\Delta \log N = \log N_{\text{cal}} - \log N_{\text{obs}}$, for each element in the sample. The typical measurement uncertainties in $\log N$ are estimated to lie within the range of 0.2 to 0.3 dex, which is shown by the dotted lines. Most residuals are found to be randomly distributed around zero, without any significant systematic trends, indicating the robustness of the adopted model assumptions.

The bottom panel of Figure~\ref{fig:enter-labe2} shows the root-mean-square (RMS) deviations of the offsets for each element. These RMS values remain below $\sim$0.3 dex when compared with the predictions and are considered to be consistent with zero. The agreement between predicted and observed abundances strongly supports the validity of the three-component fitting approach adopted in this study. Furthermore, the results provide compelling evidence that the adopted weak and main $r$-process patterns, supplemented by a main $s$-process contribution from low-mass AGB stars, are sufficient to explain the observed abundance patterns in both $s$-poor and $s$-rich stars in M22. These findings contribute valuable insight into the nucleosynthetic history of M22 and reinforce the utility of such fitting techniques for unraveling its complex chemical evolution. 

\begin{figure*}[!ht]
    \centering
    \includegraphics[width=0.49\textwidth]{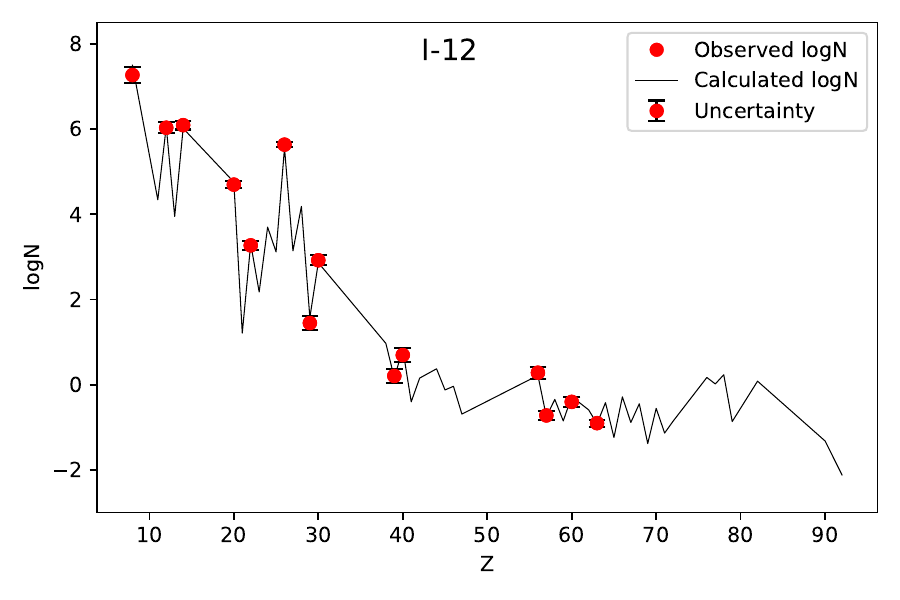}\hfill%
    \includegraphics[width=0.49\textwidth]{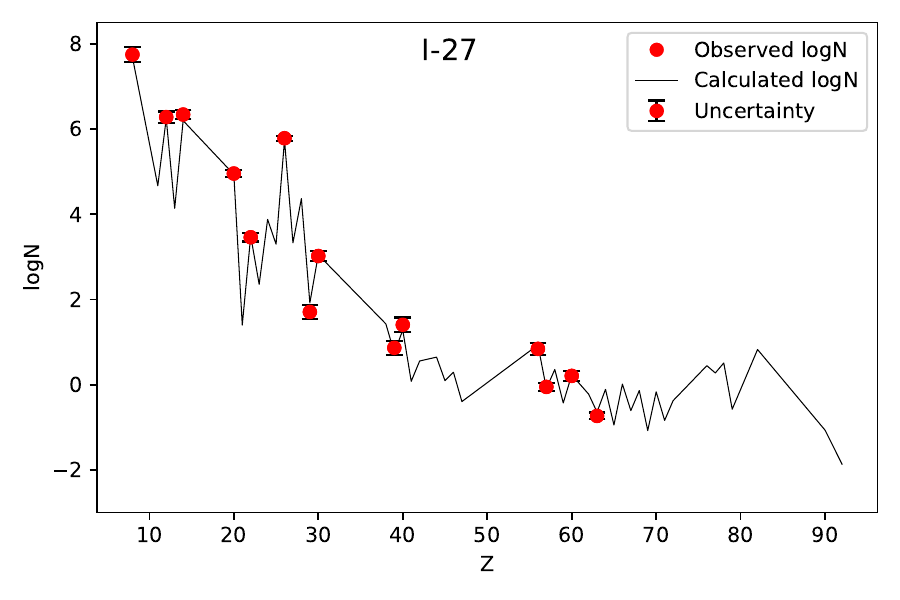}
    \includegraphics[width=0.49\textwidth]{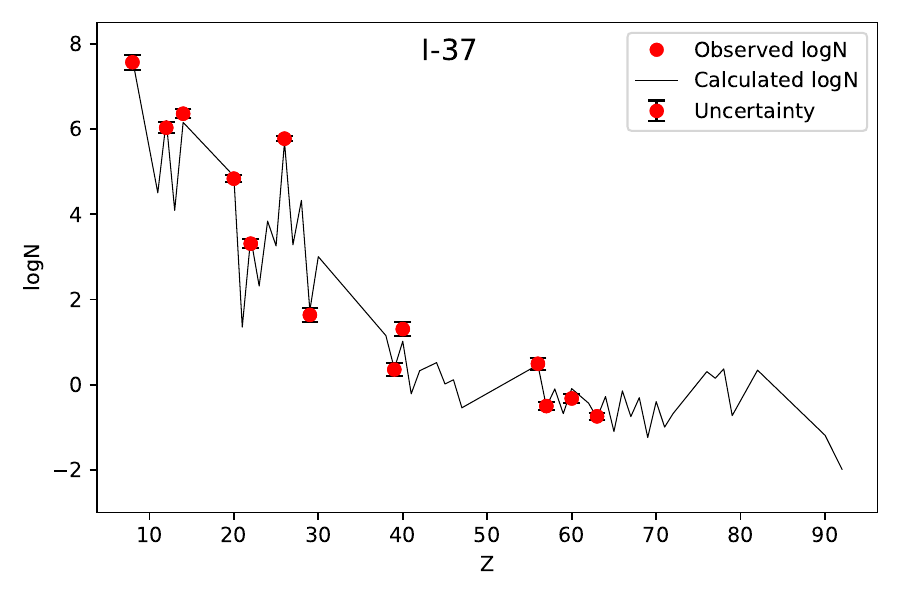}\hfill%
    \includegraphics[width=0.49\textwidth]{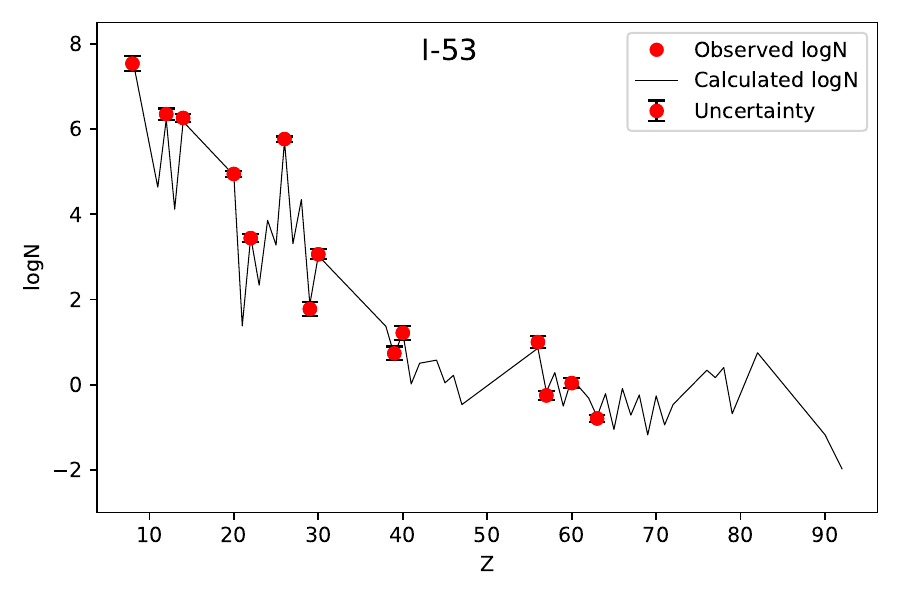}
    \includegraphics[width=0.49\textwidth]{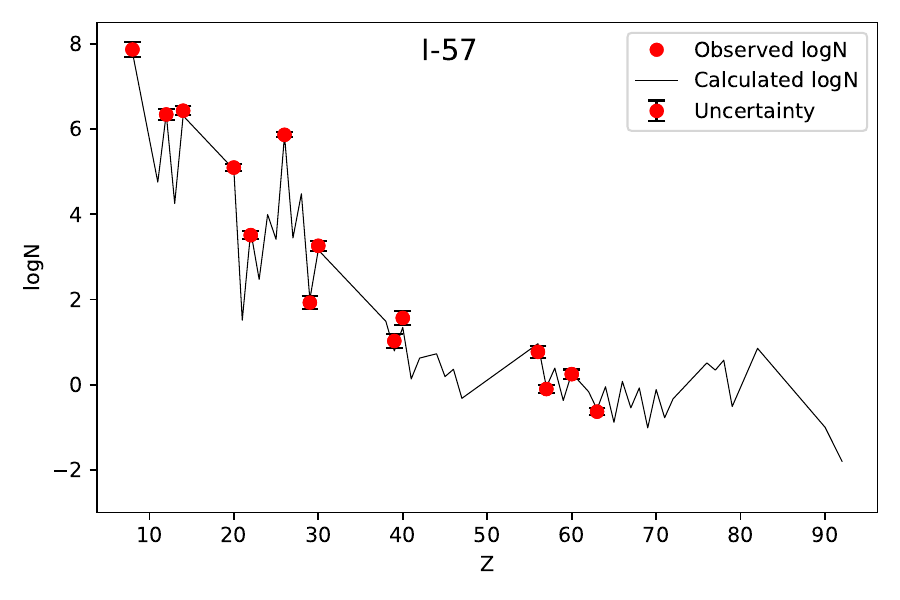}\hfill%
    \includegraphics[width=0.49\textwidth]{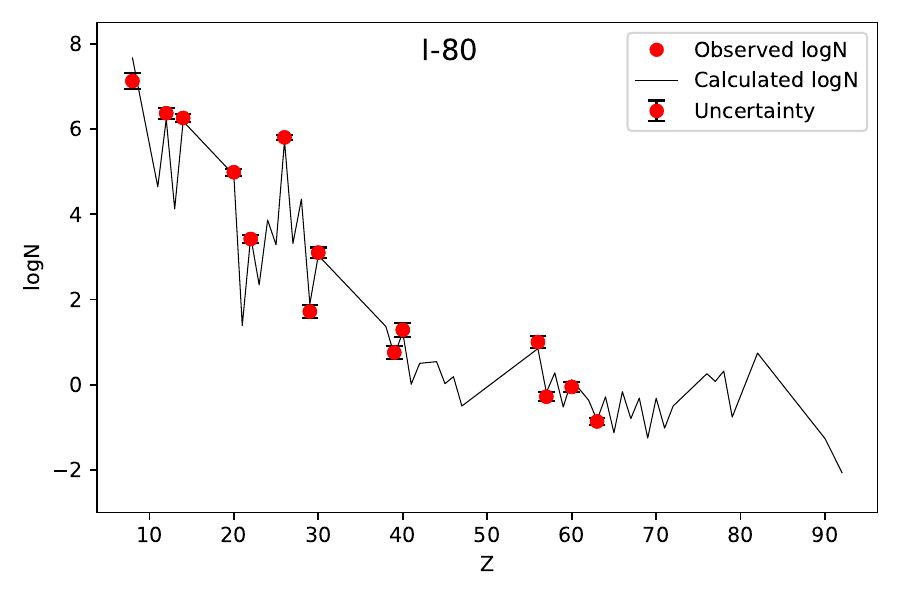}
    \includegraphics[width=0.49\textwidth]{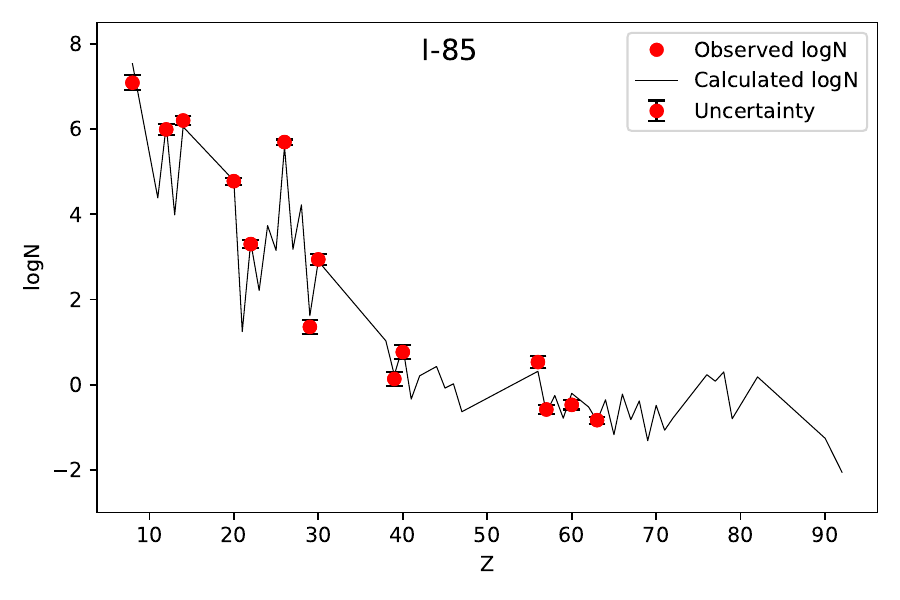}\hfill%
    \includegraphics[width=0.49\textwidth]{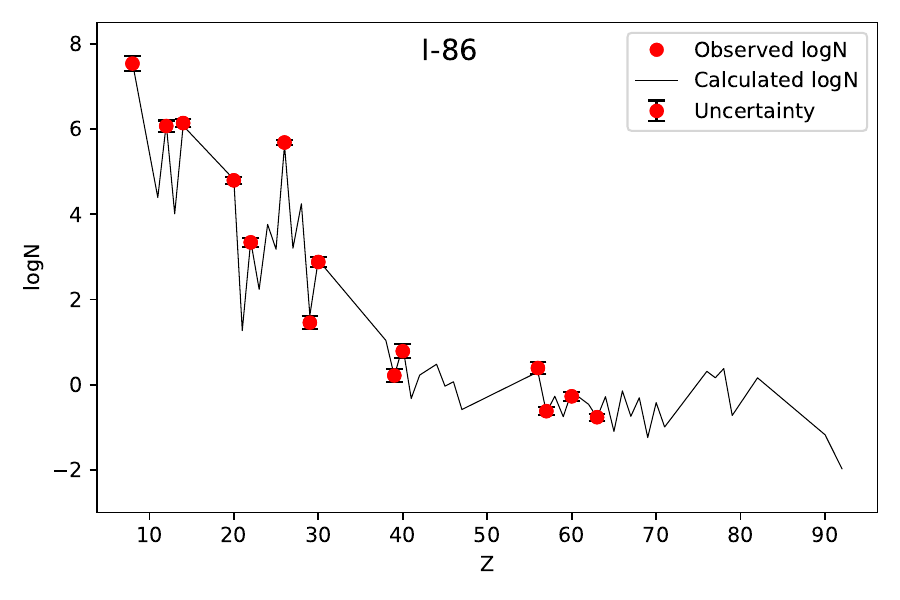}
    \caption{The calculated best-fit results. The observed elemental abundances are marked by red filled circles. The solid lines represent the best-fit results.}
    \label{fig:best-fit-1}
\end{figure*}

\addtocounter{figure}{-1}
\begin{figure*}[!ht]
    \centering
    \includegraphics[width=0.49\textwidth]{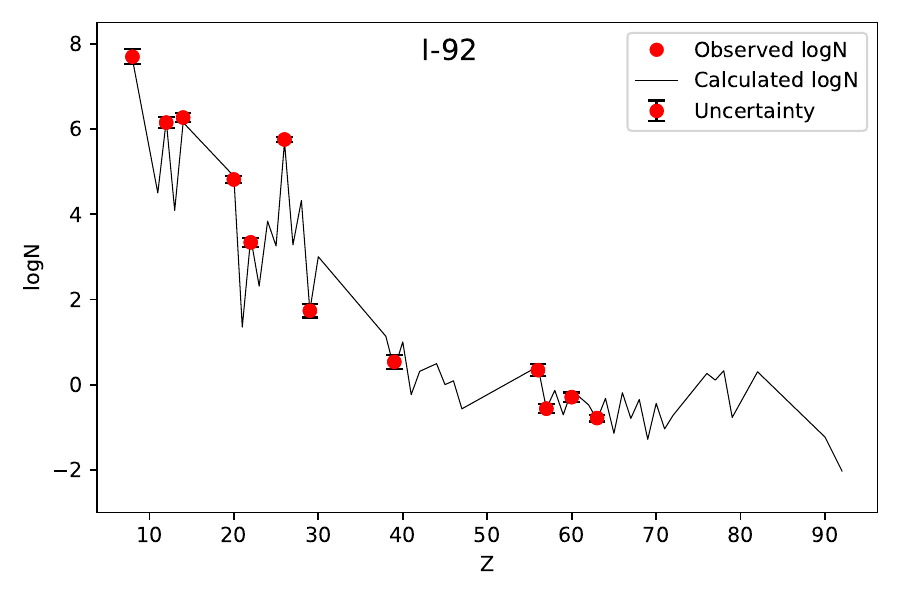}\hfill%
    \includegraphics[width=0.49\textwidth]{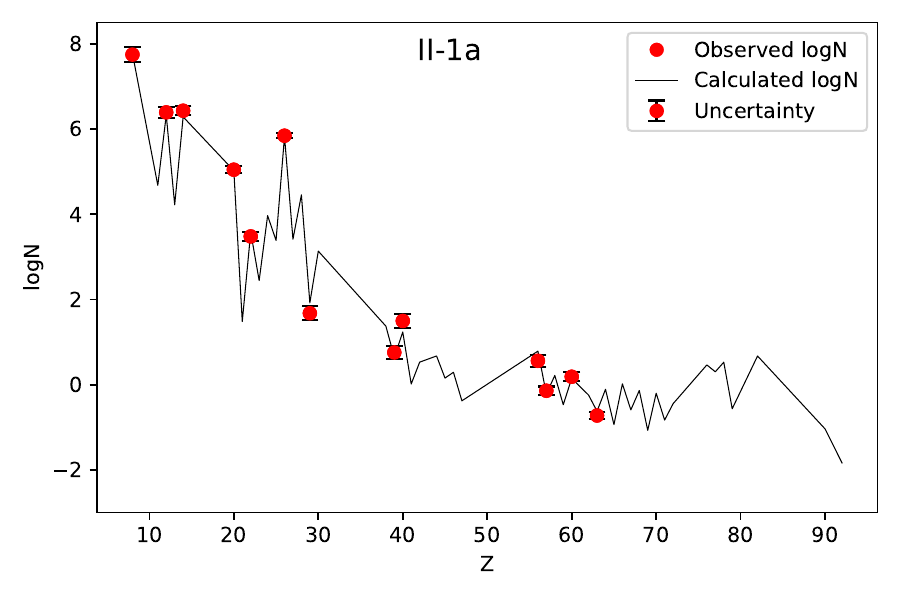}
    \includegraphics[width=0.49\textwidth]{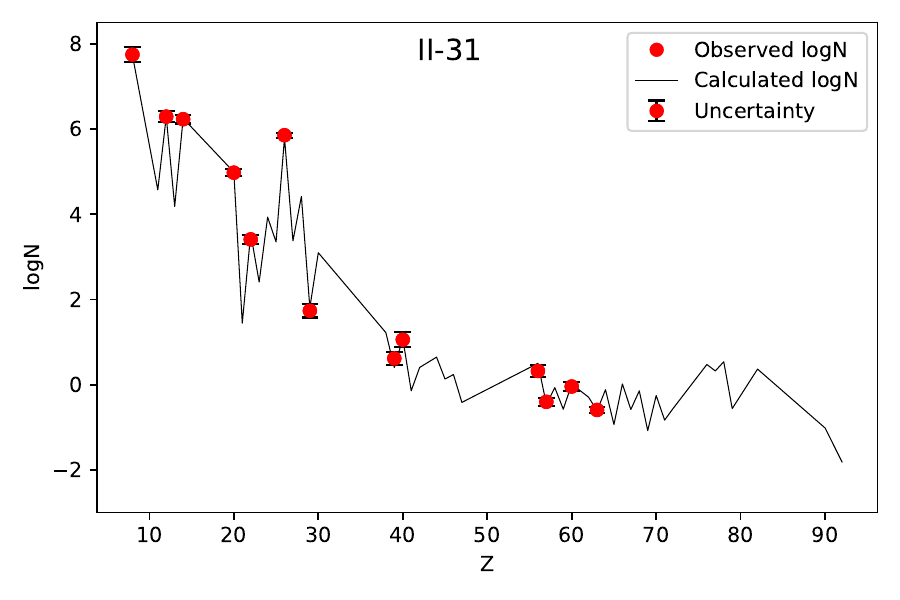}\hfill%
    \includegraphics[width=0.49\textwidth]{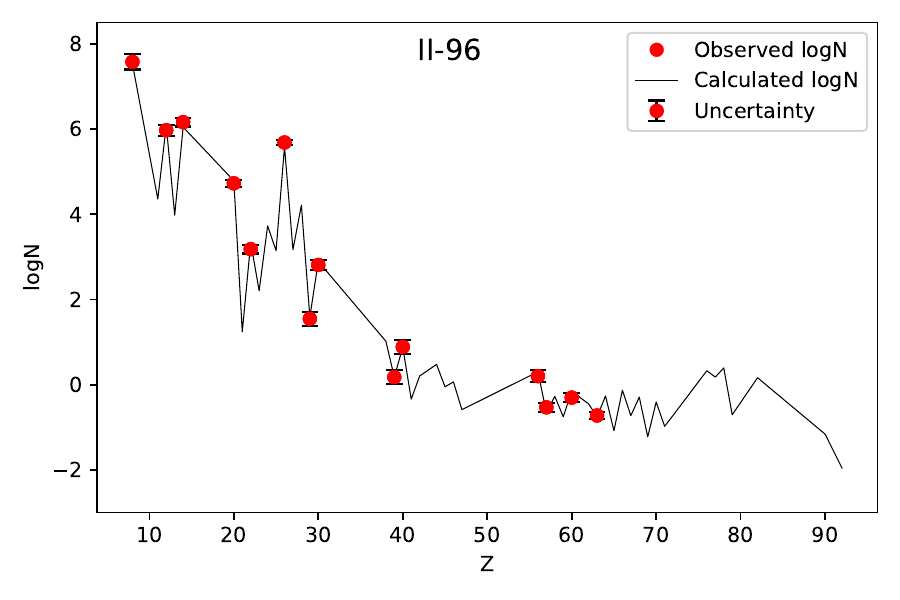}
    \includegraphics[width=0.49\textwidth]{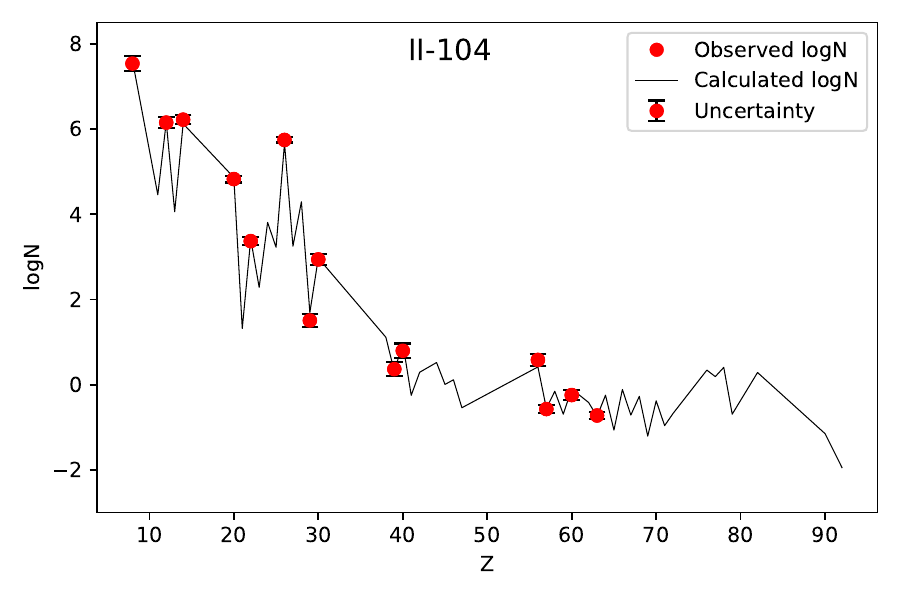}\hfill%
    \includegraphics[width=0.49\textwidth]{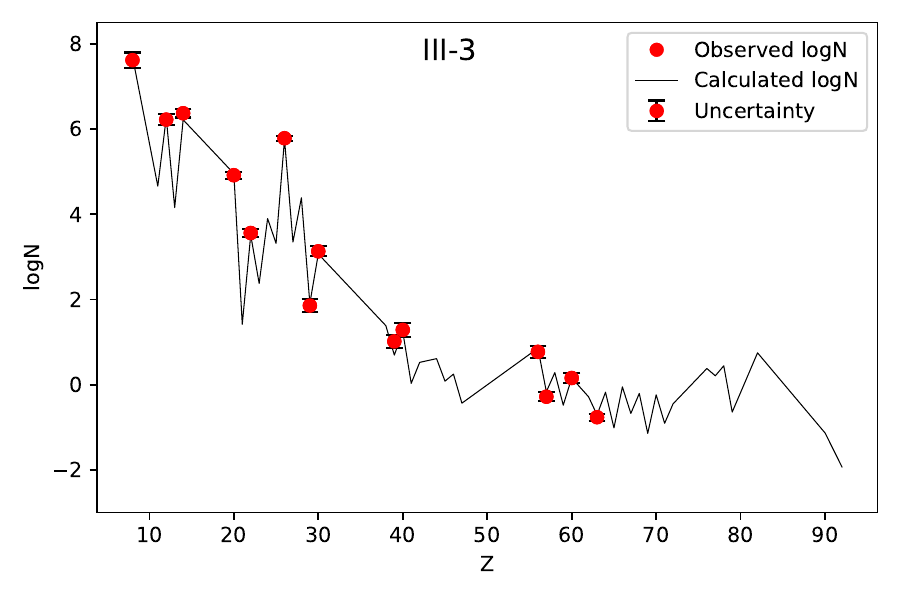}
    \includegraphics[width=0.49\textwidth]{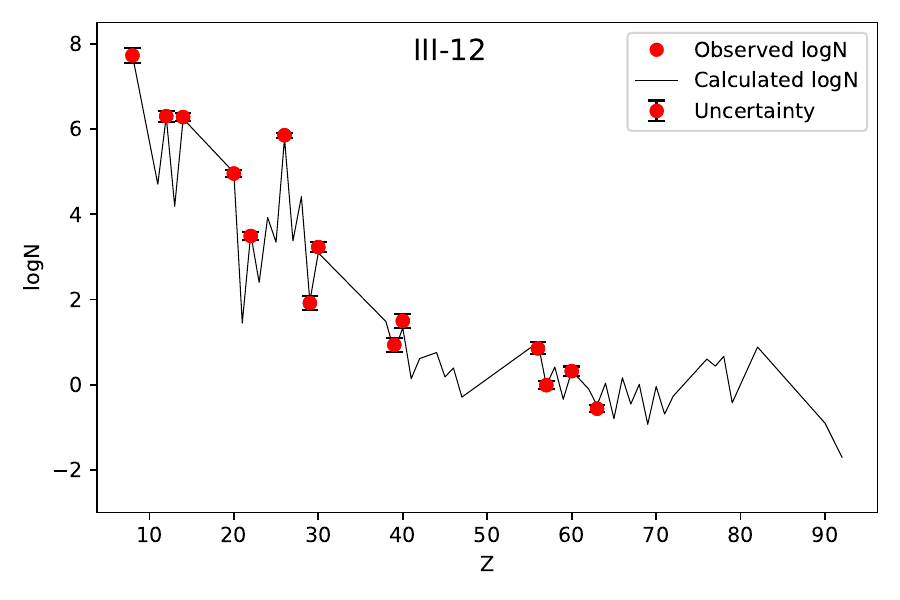}\hfill%
    \includegraphics[width=0.49\textwidth]{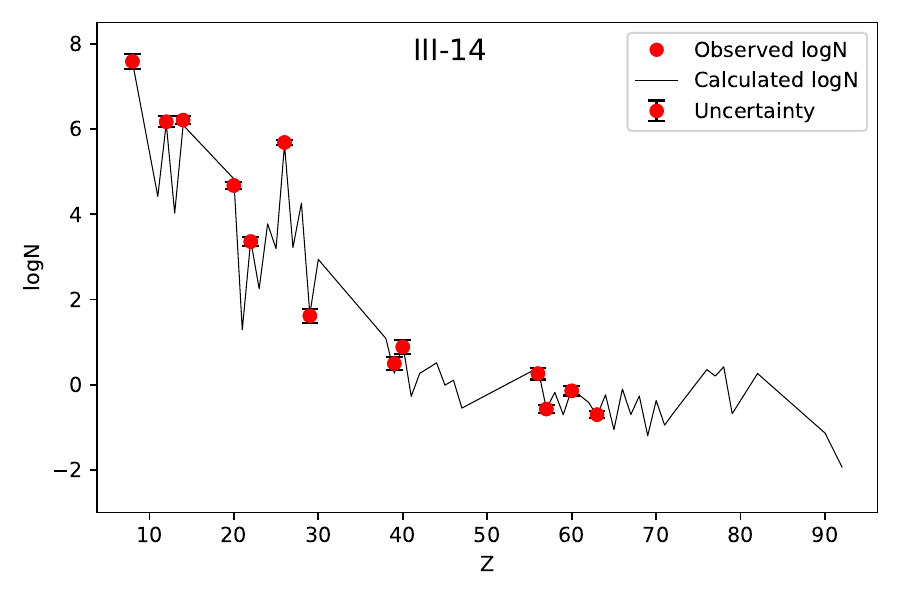}
    \caption{continued.}
\end{figure*}

\addtocounter{figure}{-1}
\begin{figure*}[!ht]
    \centering 
    \includegraphics[width=0.49\textwidth]{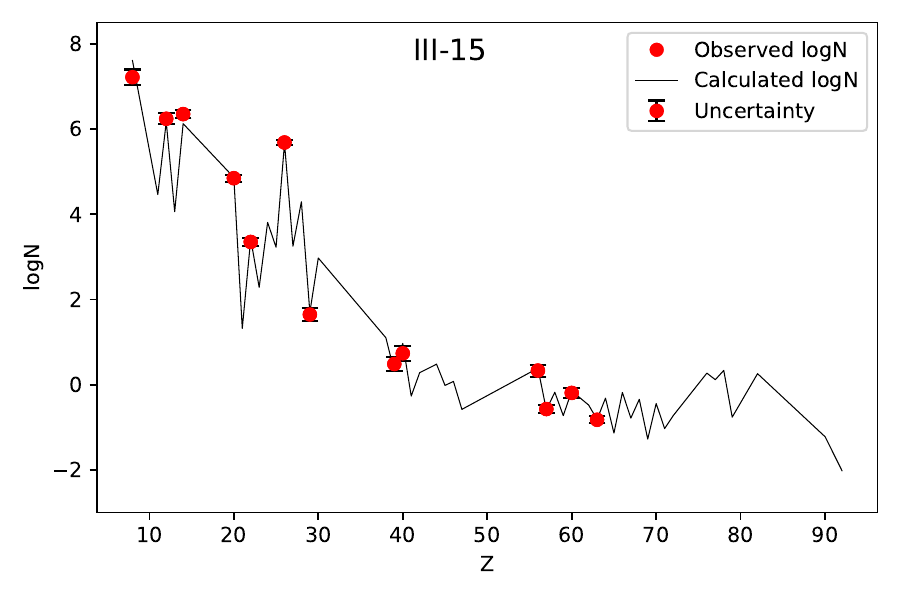}\hfill%
    \includegraphics[width=0.49\textwidth]{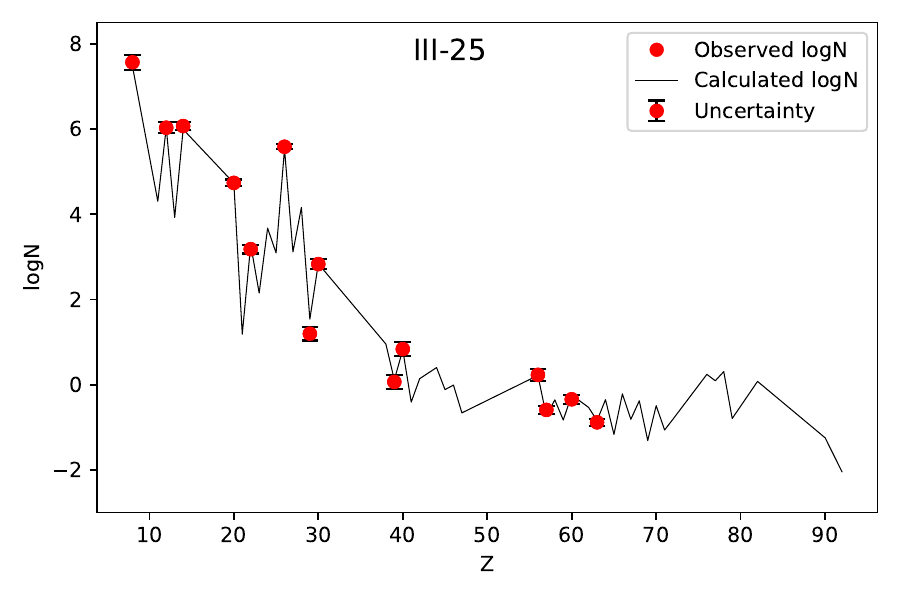}
    \includegraphics[width=0.49\textwidth]{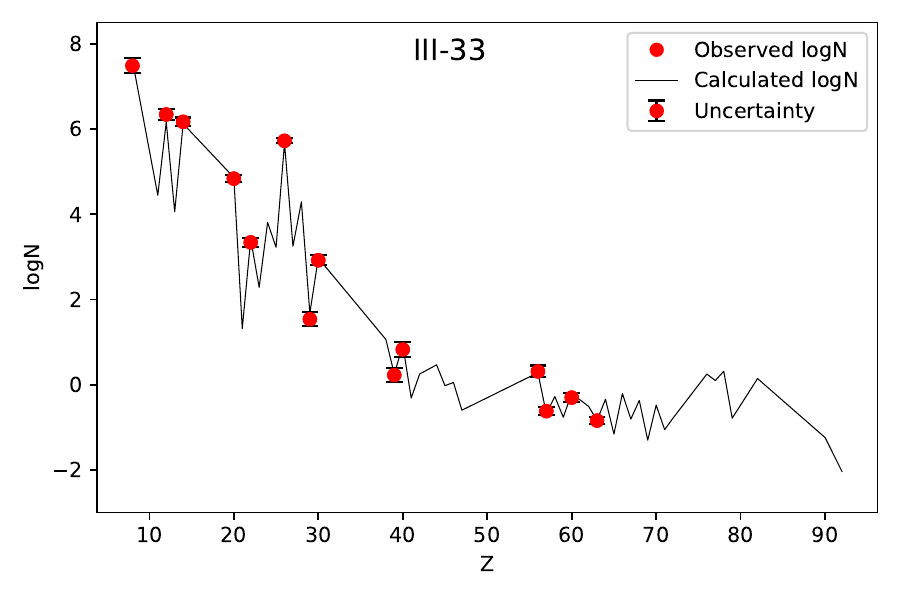}\hfill%
    \includegraphics[width=0.49\textwidth]{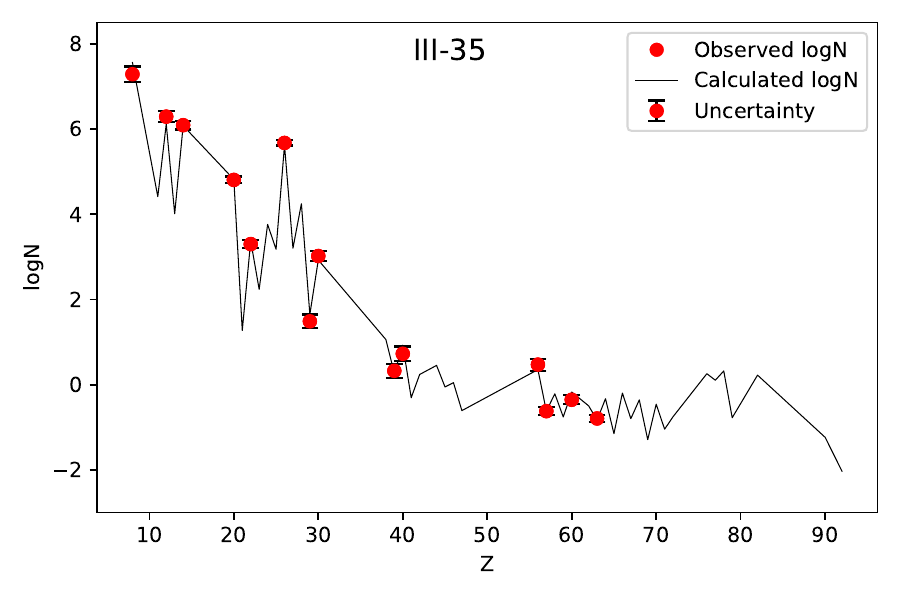}
    \includegraphics[width=0.49\textwidth]{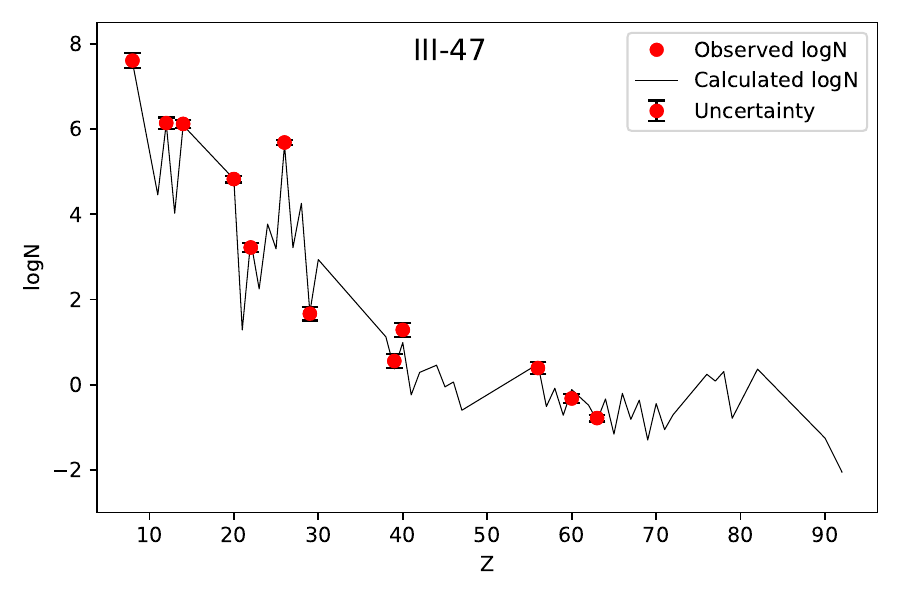}\hfill%
    \includegraphics[width=0.49\textwidth]{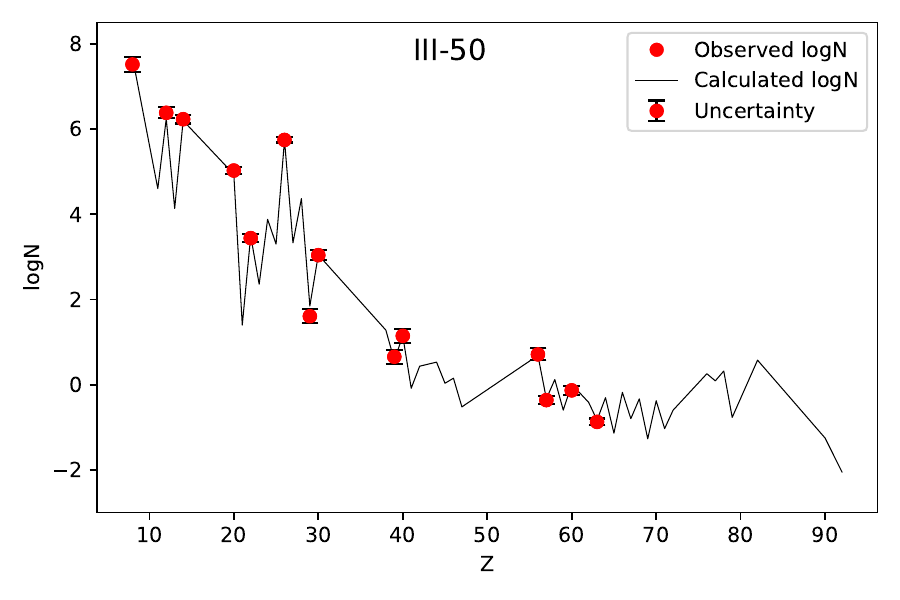}
    \includegraphics[width=0.49\textwidth]{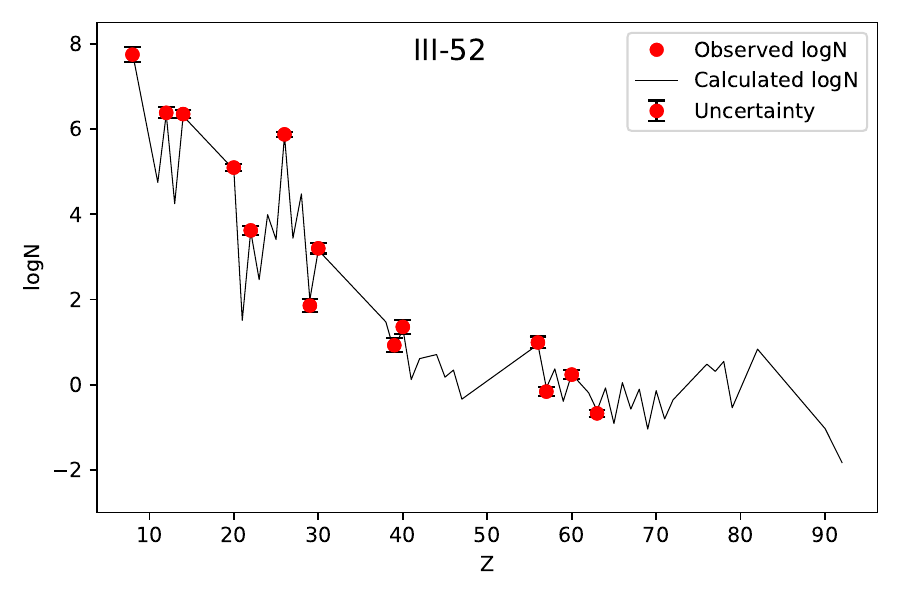}\hfill%
    \includegraphics[width=0.49\textwidth]{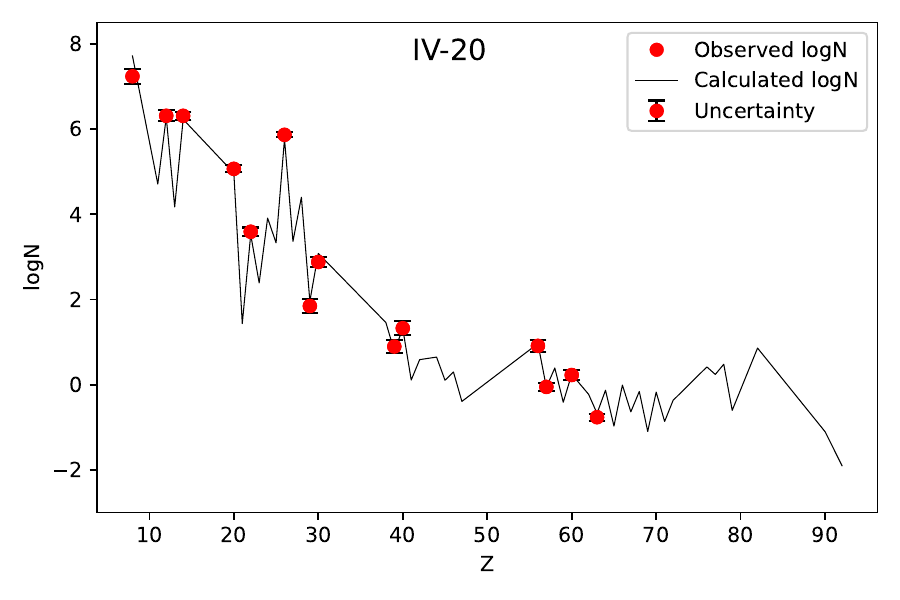}
    \caption{continued.}
\end{figure*}

\addtocounter{figure}{-1}
\begin{figure*}[!ht]
    \centering
    \includegraphics[width=0.49\textwidth]{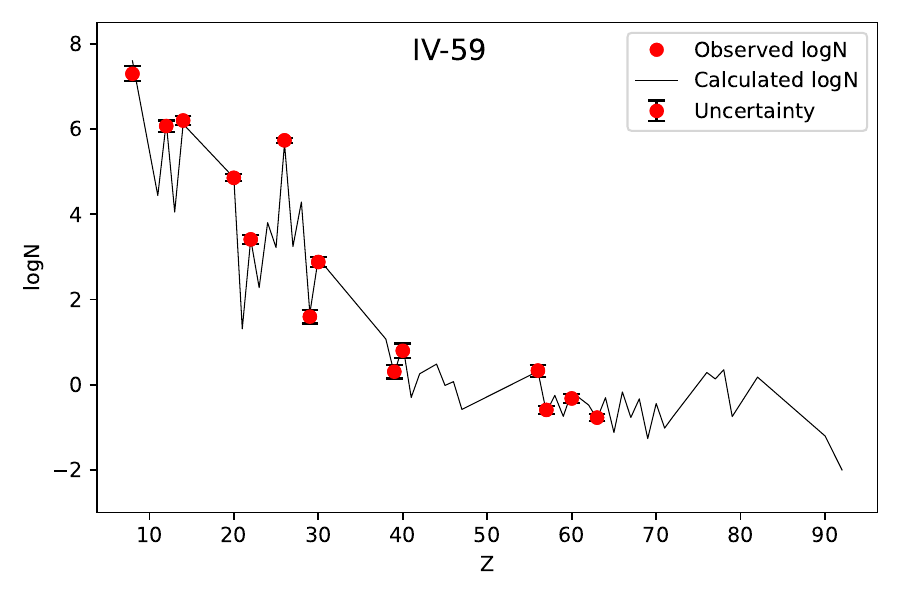}\hfill%
    \includegraphics[width=0.49\textwidth]{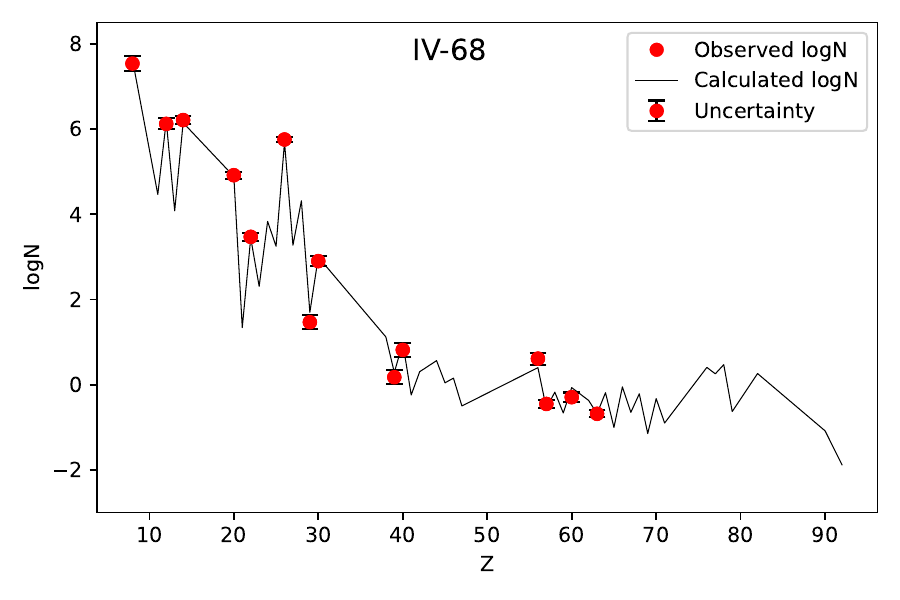}
    \includegraphics[width=0.49\textwidth]{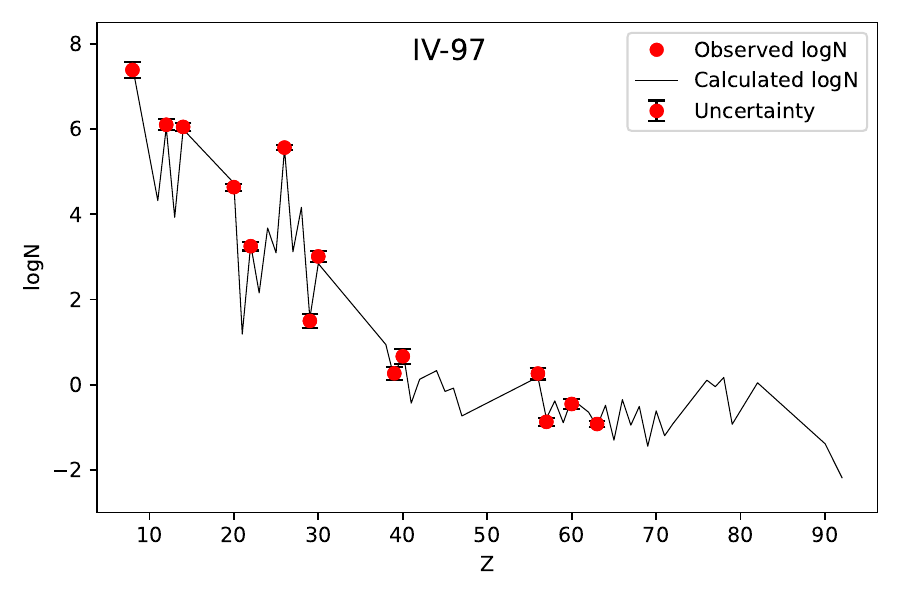}\hfill%
    \includegraphics[width=0.49\textwidth]{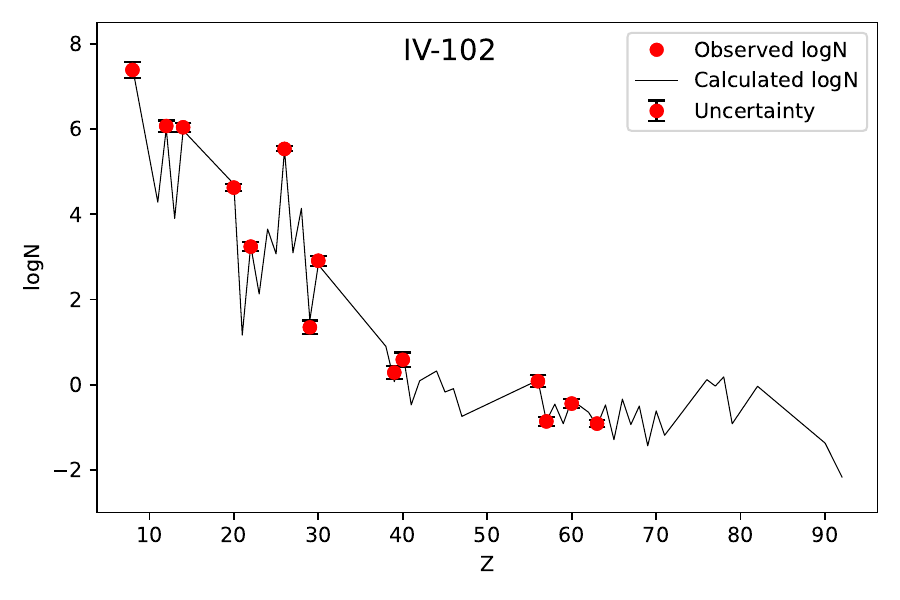}
    \includegraphics[width=0.49\textwidth]{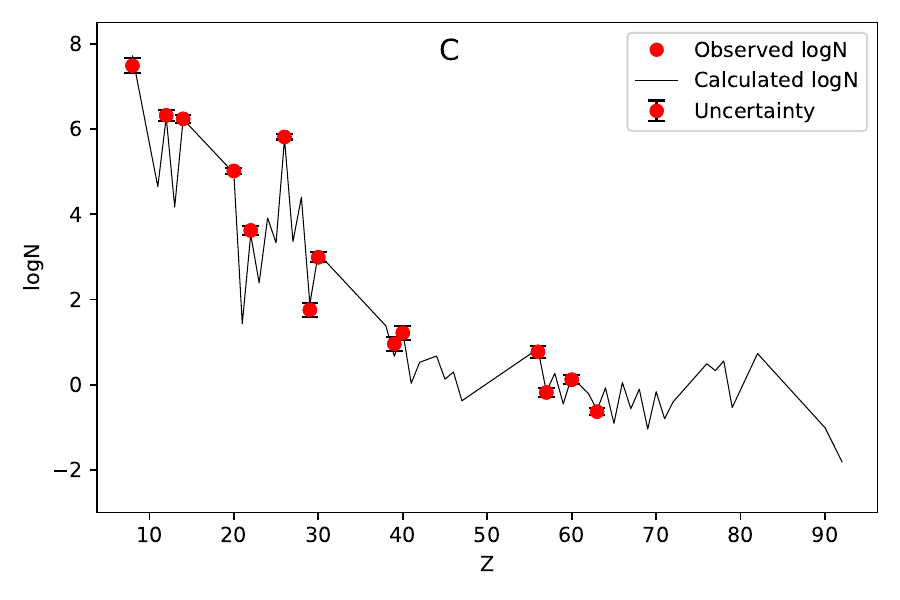}\hfill%
    \includegraphics[width=0.49\textwidth]{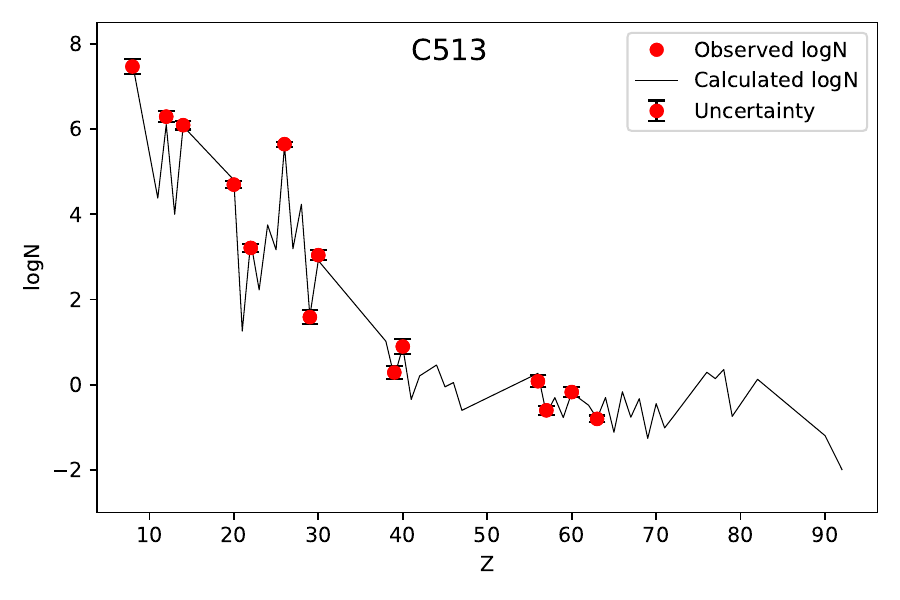}
    \caption{continued.}
    \label{fig:enter-labe1}
\end{figure*}


From Figure~\ref{fig:enter-labe1}, we also observe good agreement between the predicted and observed abundances of $s$-rich stars in M22, which implies that the origin of the $s$-process elements in M22 arises from low-mass AGB stars. 

\begin{figure}[!ht]
    \centering
    \includegraphics[width=0.8\linewidth]{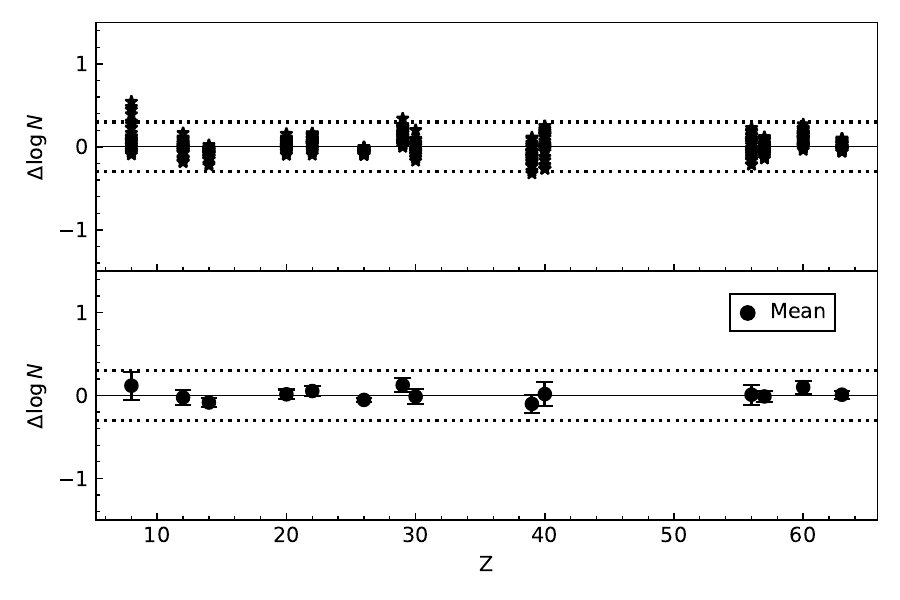}
    \caption{Top panel: Individual relative offsets ($\Delta \log N = \log N_{i,\text{cal}} - \log N_{i,\text{obs}}$) for the sample stars with respect to the predictions. Typical observational uncertainties in $\ logN$ are ~ 0.2 -0.3 dex (dotted lines). Bottom panel: The root-mean-square offsets of these elements in $\ logN$ (filled circles).}
    \label{fig:enter-labe2}
\end{figure}

\begin{figure}[!ht]
    \centering
    \includegraphics[width=0.8\linewidth]{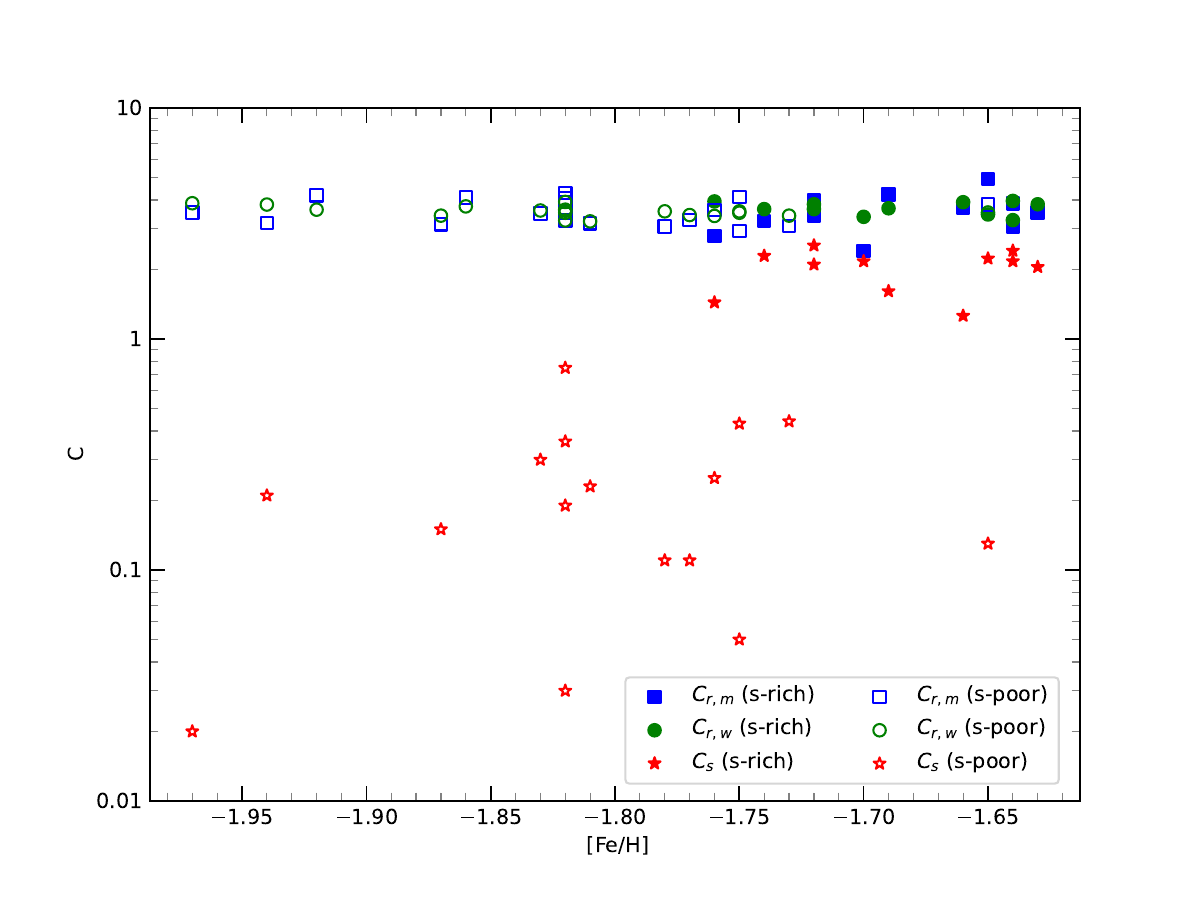}
    \caption{The component coefficients as a function of metallicity. Symbols: open squares, open circles, and open stars represent the component coefficients for the main $r$-process, weak $r$-process, and main $s$-process, respectively, in $s$-poor stars. Filled squares, filled circles, and filled stars represent the component coefficients for the main $r$-process, weak $r$-process, and main $s$-process, respectively, in $s$-rich stars.}
    \label{fig:enter-labe3}
\end{figure}

The component coefficients as a function of metallicity, illustrated in Figure~\ref{fig:enter-labe3}, provide important insights into the pollution history of M22. As shown in Figure~\ref{fig:enter-labe3}, the sample stars are clearly divided into two groups based on $C_s$: $s$-poor stars with coefficients $C_s$ smaller than 1.0 and $s$-rich stars with coefficients $C_s$ greater than 1.0. The trend of coefficients \(C_{r,m}\) and \(C_{r,w}\) remains nearly constant for the sample stars, which is distinctly different from that of $C_s$. Our results show that $C_{rw}$ remains nearly constant suggest that the elements produced by weak $r$-process have increased along with Fe throughout the pollution history of the stellar stream. This implies that both weak $r$-process elements and Fe are produced as primary elements in Type II supernovae (SNe II) and their yields have maintained a nearly constant mass fraction.

\citet{Marino2011} found that most $\alpha$ and iron-peak elements exhibit small or moderate dispersions and show no evolution with metallicity. This also suggests that both $\alpha$ and Fe-peak elements are primarily produced as primary elements in SNe II, where the weak $r$-process occurs, and their yields have remained nearly constant.

Moreover, \citet{Marino2011} observed that stars in M22 are distinctly separated into $s$-poor and $s$-rich groups based on the [La/Eu] ratio. From Figure~\ref{fig:enter-labe3}, it is evident that the $s$-component coefficients ($C_s$) increase with increasing [Fe/H]. This upward trend in $C_s$ with metallicity reflects the gradual increase in the production of main $s$-process elements relative to iron, which can be primarily attributed to the growing contribution from low-mass AGB stars. These stars, typically in the mass range of 1.5 to 3 $M_\odot$, have relatively long evolutionary timescales. At higher metallicities, these stars begin to contribute more prominently to the interstellar medium through the $s$-process, particularly via the $^{13}$C($\alpha$,n)$^{16}$O reaction \citep{Gallino1998}. The development of efficient $^{13}$C pockets at higher metallicity enhances neutron exposure, facilitating the production of heavier $s$-process elements \citep{Busso1999}. This increase also implies that the main $s$-process likely began noticeably later than the weak $r$-process and main $r$-process, which is consistent with the longer lifetimes of low-mass AGB stars. Based on the discussion above, we can infer that the $s$-rich stars observed in the M22 globular cluster represent a second generation of stellar populations, formed from molecular clouds enriched by the $s$-process yields of first-generation low-mass AGB stars. Their enhanced $s$-process abundances reflect the delayed but significant contribution  of AGB nucleosynthesis in the chemical evolution of the cluster.

\begin{table*}[t]
    \centering
    \vspace*{-\baselineskip} 
    \caption{Component Coefficients for s-rich and s-poor Stars}
    \setlength{\tabcolsep}{8pt} 
    \parbox{0.5\textwidth}{ 
        \centering
        \begin{tabular}{c c c c c c c}
            \hline
            \textbf{Star} & \textbf{[Fe/H]} & \textbf{$C_{rm}$} & \textbf{$C_{rw}$} & \textbf{$C_{s}$} & 
            \textbf{\(\chi^2\)}&
            \textbf{Type}  \\
            \hline          \\
            I-12 & -1.87 & 3.14  & 3.42 & 0.15 & 0.71 & $s$-poor \\
            I-27 & -1.72 & 3.98  & 3.65 & 2.54 & 0.71 & $s$-rich \\
            I-37 & -1.73 & 3.09  & 3.42 & 0.44 & 1.64 & $s$-poor \\
            I-53 & -1.74 & 3.24  & 3.65 & 2.29 & 0.61 & $s$-rich \\
            I-57 & -1.64 & 3.87 & 3.96 & 2.17 & 0.92 & $s$-rich \\
            I-80 & -1.7 & 2.40 & 3.38 & 2.17 & 1.83 & $s$-rich \\
            I-85 & -1.81 & 3.17 & 3.23 & 0.23 & 2.25 & $s$-poor \\
            I-86 & -1.82 & 3.91 & 3.52 & 0.03 & 0.46 & $s$-poor \\
            I-92 & -1.75 & 2.93 & 3.57 & 0.43 & 0.79 & $s$-poor \\
            II-1 & -1.66 & 3.71 & 3.91 & 1.26 & 1.23 & $s$-rich \\
            II-31 & -1.65 & 3.82 & 3.53 & 0.13 & 0.66 & $s$-poor \\
            II-96 & -1.82 & 4.06 & 3.26 & 0.00 & 1.12 & $s$-poor \\
            II-104 & -1.76 & 3.62 & 3.41 & 0.25 & 0.83 & $s$-poor \\
            III-3 & -1.72 & 3.42 & 3.83 & 2.10 & 0.91 & $s$-rich \\
            III-12 & -1.65 & 4.92 & 3.46 & 2.23 & 0.63 & $s$-rich \\
            III-14 & -1.82 & 4.28 & 3.63 & 0.19 & 0.86 & $s$-poor \\
            III-15 & -1.82 & 3.52 & 3.92 & 0.36 & 1.30 & $s$-poor \\
            III-25 & -1.92 & 4.19 & 3.63 & 0.00 & 0.96 & $s$-poor \\
            III-33 & -1.78 & 3.07 & 3.57 & 0.11 & 0.62 & $s$-poor \\
            III-35 & -1.83 & 3.50 & 3.60 & 0.30 & 1.1 & $s$-poor \\
            III-47 & -1.82 & 3.26 & 3.59 & 0.75 & 1.15 & $s$-poor \\
            III-50 & -1.76 & 2.79 & 3.94 & 1.44 & 0.75 & $s$-rich \\
            III-52 & -1.63 & 3.53 & 3.83 & 2.05 & 0.38 & $s$-rich \\
            IV-20 & -1.64 & 3.05 & 3.27 & 2.41 & 1.58 & $s$-rich \\
            IV-59 & -1.77 & 3.27 & 3.44 & 0.11 & 0.85 & $s$-poor \\
            IV-68 & -1.75 & 4.12 & 3.52 & 0.05 & 1.26 & $s$-poor \\
            IV-97 & -1.94 & 3.18 & 3.82 & 0.21 & 0.75 & $s$-poor \\
            IV-102 & -1.97 & 3.53 & 3.87 & 0.02 & 0.56 & $s$-poor \\
            C & -1.69 & 4.23 & 3.68 & 1.61 & 0.82 & $s$-rich \\
            C513 & -1.86 & 4.09 & 3.75 & 0.00 & 0.99 & $s$-poor \\
            
            \hline
        \end{tabular}
         \label{tab:1}
    }
\end{table*}
\section{Conclusion}
\label{sec:4}
In globular clusters, the vast majority of chemical evolution and nucleosynthetic information is encoded in the elemental abundances of stars exhibiting a range of metallicities. In this context, the chemical abundances of metal-poor stars in M22 serve as invaluable data for constraining theoretical models of both the $r$ and $s$-process nucleosynthesis. By analyzing the abundance patterns in these stars, we can gain critical insights into the complex nucleosynthetic history and the diverse astrophysical processes that contributed to their chemical composition. Our findings can be summarized as follows:
\begin{enumerate}
    \item The abundances of most $s$-poor stars in M22, including $\alpha$, Fe-peak, and neutron-capture elements, are best reproduced by a combination of weak and main $r$-process patterns. The coefficients $C_{r,w}$ and $C_{r,m}$ exhibit a nearly constant trend across the sample stars, including $s$-rich stars. Despite the low metallicities of the sample stars, the ratios of weak and main $r$-process components have reached values comparable to those observed in the solar system, i.e., $C_{r,w} \approx C_{r,m}$.

   \item For $s$-rich stars in M22, the most plausible origin of the $s$-process elements is pollution by low-mass AGB stars. 

\begin{itemize}\item[(i)] The $s$-process elements in $s$-rich stars are primarily a result of pollution from AGB stars, with their $r$-component coefficients remaining comparable to those of $s$-poor stars. \item[(ii)] The trends of component coefficients for the $s$-process and $r$-process in M22 exhibit noticeable differences. For the sample stars, the increasing trend in $C_{s}$ with increasing [Fe/H] indicates a gradual rise in the contribution of the main $s$-process. This trend can be attributed to the longer lifetime of low-mass AGB stars.
\end{itemize}
\end{enumerate}

Clearly, it is crucial for future studies to determine the specific evolutionary scenario of M22. Further theoretical and observational investigations will enhance our understanding of the r-process at low metallicity and provide insight into the history of neutron-capture element enrichment in globular clusters.

\section*{Funding}
This study is supported by the National Key Basic R\&D Program of China No. 2024YFA1611903, the National Natural Science Foundation of China under grant No. 12173013, the project of Hebei provincial department of science and technology under the grant number 226Z7604G.

\bibliographystyle{Frontiers-Harvard}
\bibliography{test}
\end{document}